\documentclass[runningheads]{svmult}
\usepackage{makeidx}   
\usepackage{graphicx}  
\usepackage{subeqnar}  
\usepackage{multicol}  
\usepackage{cropmark} 
\usepackage{physprbb}  
\begin{document}
\title*{Neutrino physics from cosmological observations}
\toctitle{Neutrino physics from cosmological observations}
%
%
\titlerunning{Neutrino physics}
%
\author{Steen Hannestad}
\authorrunning{Steen Hannestad}
%
%
\institute{Department of Physics, University of Southern Denmark
     Campusvej 55,  \\ DK-5230 Odense M, Denmark}

\maketitle              

\begin{abstract}
Cosmological observations have in the past few years become an
increasingly powerful method for determining parameters within
the neutrino sector, such as the presence of sterile states and the
mass of neutrinos. I review the current status of the field in light
of recent measurements of the cosmic microwave background
by the WMAP collaboration, as well as current large scale galaxy surveys.
\end{abstract}

\section{Introduction}

Neutrinos exist in equilibrium with the electromagnetic plasma
in the early universe, until a temperature of a few MeV. At this
point the weak interactions freeze out and neutrinos decouple from
the plasma. Shortly after this event, the temperature of the plasma
falls below the electron mass, and electrons and positrons annihilate,
dumping their entropy into the photon gas. This heats the photon
gas while having no effect on neutrinos, and the end result is
that the photon temperature is larger than the neutrino temperature
by the factor $(11/4)^{1/3} \simeq 1.40$. Since the present day
photon temperature has been measured with great accuracy to be
2.728 K, the neutrino temperature is known to be 1.95 K, or
about $2 \times 10^{-4}$ eV. Since the heaviest neutrino has
a mass of at least about 0.04 eV it must at present be extremely
non-relativistic and therefore acts as dark matter.
The contribution of a single neutrino species of mass $m_\nu$
to the present day matter density can be written as
\cite{Hannestad:1995rs,Dolgov:1997mb,Mangano:2001iu}
\begin{equation}
\Omega_\nu h^2 = \frac{m_\nu}{92.5 {\rm eV}},
\end{equation}
so that for a neutrino mass of about 30 eV, neutrinos will make
up all of the dark matter.
However, this would have disastrous consequences for structure
formation in the universe, because neutrinos of eV mass have very
large free streaming lengths and would erase structure in the
neutrino density on 
scales smaller than $l_{\rm fs} \simeq 1 \,\, 
m_{\nu,{\rm eV}}^{-1} \,\,\, {\rm Gpc}$
completely.
This leads to an overall suppression of matter fluctuations at
small scales, an effect which is potentially observable.


\subsection{Absolute value of neutrino masses}

The absolute value of neutrino masses are very difficult to measure
experimentally. On the other hand, mass differences between neutrino
mass eigenstates, $(m_1,m_2,m_3)$, 
can be measured in neutrino oscillation experiments.
Observations of atmospheric neutrinos suggest a squared mass 
difference of $\delta m^2 \simeq 3 \times 10^{-3}$ eV$^2$
\cite{Fukuda:2000np,Fornengo:2000sr,Maltoni:2002ni}. While there are still
several viable solutions to the solar neutrino problem from solar
neutrino observations alone,
the large mixing angle (LMA) solution gives by far the best fit with
$\delta m^2 \simeq 5 \times 10^{-5}$ eV$^2$ \cite{sno,Bahcall:2002hv}. 
Recently the KamLAND reactor neutrino experiment has announced a positive
detection of neutrino oscillations indicating that the LMA solution
is indeed correct \cite{kamland}.

In the simplest case where neutrino masses are
hierarchical these results suggest that $m_1 \sim 0$, $m_2 \sim 
\delta m_{\rm solar}$, and $m_3 \sim \delta m_{\rm atmospheric}$.
If the hierarchy is inverted 
\cite{Kostelecky:1993dm,Fuller:1995tz,Caldwell:1995vi,Bilenky:1996cb,King:2000ce,He:2002rv}
one instead finds
$m_3 \sim 0$, $m_2 \sim \delta m_{\rm atmospheric}$, and 
$m_1 \sim \delta m_{\rm atmospheric}$.
However, it is also possible that neutrino
masses are degenerate
\cite{Ioannisian:1994nx,Bamert:vc,Mohapatra:1994bg,Minakata:1996vs,Vissani:1997pa,Minakata:1997ja,Ellis:1999my,Casas:1999tp,Casas:1999ac,Ma:1999xq,Adhikari:2000as}, 
$m_1 \sim m_2 \sim m_3 \gg \delta m_{\rm atmospheric}$, 
in which case oscillation experiments are not
useful for determining the absolute mass scale.

Experiments which rely on kinematical effects of the neutrino mass
offer the strongest probe of this overall mass scale. Tritium decay
measurements have been able to put an upper limit on the electron
neutrino mass of 2.2 eV (95\% conf.) \cite{Bonn:tw}.
However, cosmology at present yields an even stronger limit which
is also based on the kinematics of neutrino mass.
As discussed before any structure in the neutrino density below
the free-streaming scale is erased and therefore the presence
of a non-zero neutrino mass suppresses the matter power spectrum
at small scales relative to large scale, roughly by
$\Delta P/P \sim -8 \Omega_\nu/\Omega_m$ \cite{Hu:1997mj}.

This power spectrum suppression allows for a determination of the
neutrino mass from measurements of the matter power spectrum on
large scales, as well as the spectrum of CMB fluctuations. 
This matter spectrum is related to the galaxy correlation
spectrum measured in large scale structure (LSS) surveys via the
bias parameter, $b^2(k) \equiv P_g(k)/P_m(k)$.
Such analyses have been performed several times before
\cite{Croft:1999mm,Fukugita:1999as}, most recently
using data from the 2dFGRS galaxy survey 
\cite{Elgaroy:2002bi,Hannestad:2002xv,Lewis:2002ah}. 
These investigations found mass limits of 1.5-3 eV, depending on
assumptions about the cosmological parameter space.

In a seminal paper it was calculated by Eisenstein, Hu and Tegmark
that future CMB and LSS experiments could push the bound on the
sum of neutrino masses down to about 0.3 eV \cite{Hu:1997mj}.
The prospects for an absolute
neutrino mass determination was discussed in further detail in
Ref.~\cite{Hannestad:2002cn} where it was found that in fact the 
upper bound could
be pushed to 0.12 eV (95\% conf.) using data from the Sloan
Digital Sky Survey and the upcoming
Planck satellite.

More recently the new WMAP data, in conjunction with large scale structure
data from 2dFGRS has been used to put an upper bound on the
sum of all neutrino species of $\sum m_\nu \leq 0.7$ eV (95\% conf.)
\cite{map2}.

However, the exact value of this upper bound depends strongly
on priors on other cosmological parameters, mainly $H_0$.
In the present paper we calculate the upper bound on $\sum m_\nu$
from present cosmological data, with an emphasis on studying 
how the bound depends on the data set chosen.


In addition to their contribution to the cosmological mass density
neutrinos also contribute to the cosmological energy density
around the epoch of recombination. 
Neutrinos which have mass smaller than roughly $3T_{\rm rec}$, where
$T_{\rm rec} \simeq 0.3$ eV is the temperature of recombination, will
act as fully relativistic particles when it comes to CMB and large
scale structure.

In the standard model there are three light neutrino species with 
this property. However, these particles are not necessarily in
an equilibrium Fermi-Dirac distribution with zero chemical potential.
It is well known that the universe contains a non-zero baryon
asymmetry of the order 
$\eta = \frac{n_B - n_{\bar{B}}}{n_\gamma} \sim 10^{-10}$. 
A neutrino asymmetry of similar
magnitude would have no impact on cosmology during CMB and LSS
formation, but since the neutrino asymmetry is not directly 
observable it could in principle be much larger than the
baryon asymmetry. Such a neutrino
asymmetry would effectively show up as extra relativistic energy
density in the CMB and LSS power spectra.

Another possibility for extra relativistic energy is that 
there are additional light species beyond the standard model
which have decoupled early (such as the graviton or the gravitino). 

From the perspective of late time evolution at $T \leq 1$ MeV 
it is customary 
to parametrize any such additional energy density in terms of
$N_\nu$ \cite{Steigman:kc}, 
the equivalent number of neutrino species. In Section III
we discuss bounds on $N_\nu$ from the present WMAP and 2dFGRS data,
combined with additional information on other cosmological
parameters from the Hubble HST key project and the Supernova
Cosmology Project.

However, as will be discussed later, a non-zero neutrino chemical
potential can have an effect on big bang nucleosynthesis which is profoundly
different from simple relativistic energy density
if it is located in the electron neutrino sector.

Another important point is that any entropy production which takes place
after BBN, but prior to CMB formation will only be detectable via
CMB and LSS observations. One such example is the decay of a hypothetical
long-lived massive particle at temperatures
below roughly 0.01 MeV.


\section{Likelihood analysis and data sets}

The extraction of cosmological parameters from cosmological data
is a difficult process 
since for both CMB and LSS the power spectra depend on a plethora
of different parameters.
Furthermore, since the CMB and matter power spectra
depend on many different parameters one might
worry that an analysis which is too restricted in parameter space 
could give spuriously strong limits on a given parameter.

The most recent cosmological data is in excellent agreement with 
a flat $\Lambda$CDM model, the only non-standard feature being
the apparently very high optical depth to reionization. Therefore
the natural benchmark against which non-standard neutrino physics can
be tested is a model with the following free parameters:
$\Omega_m$, the matter density,
the curvature parameter, $\Omega_b$, the baryon density, $H_0$, the
Hubble parameter, $n_s$, the scalar spectral index of the primordial
fluctuation spectrum, $\tau$, the optical depth to reionization,
$Q$, the normalization of the CMB power spectrum, $b$, the 
bias parameter, and finally the two parameters related to neutrino physics,
$\Omega_\nu h^2$ and $N_\nu$.
The analysis can be restricted to geometrically flat models, i.e.\
$\Omega = \Omega_m + \Omega_\Lambda = 1$.
For the purpose of actual power spectrum calculations, the CMBFAST
package \cite{CMBFAST} can be used.

\subsection{LSS data}

At present, by far the largest survey available
is the 2dFGRS \cite{2dFGRS} of which about 147,000 galaxies have so far been
analyzed. Tegmark, Hamilton and Xu \cite{THX} have calculated a power
spectrum, $P(k)$, from this data, which we use in the present work.
The 2dFGRS data extends to very small scales where there are large
effects of non-linearity. Since we only calculate linear power
spectra, we use (in accordance with standard procedure) only data on
scales larger than $k = 0.2 h \,\, {\rm Mpc}^{-1}$, where effects of
non-linearity should be minimal \cite{Hannestad:2002cn}. 
Making this cut reduces the number of power spectrum data points to 18.

\subsection{CMB data}

The CMB temperature
fluctuations are conveniently described in terms of the
spherical harmonics power spectrum
$C_l \equiv \langle |a_{lm}|^2 \rangle$,
where
$\frac{\Delta T}{T} (\theta,\phi) = \sum_{lm} a_{lm}Y_{lm}(\theta,\phi)$.
Since Thomson scattering polarizes light there are additional powerspectra
coming from the polarization anisotropies. The polarization can be
divided into a curl-free $(E)$ and a curl $(B)$ component, yielding
four independent power spectra: $C_{T,l}, C_{E,l}, C_{B,l}$ and 
the temperature $E$-polarization cross-correlation $C_{TE,l}$.

The WMAP experiment have reported data only on $C_{T,l}$ and $C_{TE,l}$,
as described in Ref.~\cite{map1,map2,map3,map4,map5}

We have performed the likelihood analysis using the prescription
given by the WMAP collaboration which includes the correlation
between different $C_l$'s \cite{map1,map2,map3,map4,map5}. Foreground contamination has
already been subtracted from their published data.

In parts of the data analysis we also add other CMB data from
the compilation by Wang {\it et al.} \cite{wang3}
which includes data at high $l$.
Altogether this data set has 28 data points.


\section{Numerical results}

\subsection{Neutrino masses}

The analysis presented here was originally published in Ref.~\cite{steen03},
and more details can be found there.

We have calculated $\chi^2$ as a function of neutrino mass while
marginalizing over all other cosmological parameters. This has been
done using the data sets described above. In the first case we have
calculated the constraint using the WMAP $C_{T,l}$ combined with
the 2dFGRS data, and in the second case we have added the polarization
measurement from WMAP.
Finally we have added the additional constraint from the
HST key project and the Supernova Cosmology Project.
It should also be noted that when constraining the neutrino mass
it has in all cases been assumed that $N_\nu$ is equal to the
standard model value of 3.04. Later we relax this
condition in order to study the LSND bound.

The result is shown in Fig.~1. As can be seen from the figure
the 95\% confidence upper limit on the sum of neutrino masses is
$\sum m_\nu \leq 1.01$ eV (95\% conf.) using the case with priors.
This value is completely consistent with what is found in
Ref.~\cite{el03} where simple Gaussian priors from WMAP were
added to the 2dFGRS data analysis.
For the three cases studied we find the following limits:
\begin{tabbing}
$\sum m_\nu < 1.01$ eV \hspace*{0.8cm} \= for WMAP+Wang+2dFGRS+HST+SN-Ia \\
$\sum m_\nu < 1.20$ eV \> for WMAP+Wang+2dFGRS \\
$\sum m_\nu < 2.12$ eV \> for WMAP+2dFGRS 
\end{tabbing}

In the middle panel of Fig.~1 we show the best fit value of $H_0$
for a given $\Omega_\nu h^2$. It is clear that an increasing value
of $\sum m_\nu$ can be compensated by a decrease in $H_0$. Even though
the data yields a strong constraint on $\Omega_m h^2$ there is no
independent constraint on $\Omega_m$ in itself. Therefore, an decreasing
$H_0$ leads to an increasing $\Omega_m$. This can be seen in the bottom
panel of Fig.~1.

When the HST prior on $H_0$ is relaxed a higher value of $\sum m_\nu$
is allowed, in the case with only WMAP and 2dFGRS data the upper bound
is $\Omega_\nu h^2 \leq 0.023$ (95\% conf.), corresponding to a
neutrino mass of 0.71 eV for each of the three neutrinos.

This effect was also found by Elgar{\o}y and Lahav \cite{el03} in 
their analysis of the effects of priors on the determination 
of $\sum m_\nu$.

However, as can also be seen from the figure, the addition of high-$l$
CMB data from the Want {\it et al.} compilation also shrinks the
allowed range of $\sum m_\nu$ significantly. The reason is that
there is a significant overlap of the scales probed by high-$l$ CMB
experiments and the 2dFGRS survey. Therefore, even though we use bias
as a free fitting parameter, it is strongly constrained by the fact
that the CMB and 2dFGRS data essentially cover much of the same
range in $k$-space.

It should be noted that Elgar{\o}y and Lahav \cite{el03} find that bias
does not play any role in determining the bound on $\sum m_\nu$. At first
this seems to contradict the discussion here, and also what was found
from a Fisher matrix analysis in Ref.~\cite{Hannestad:2002xv}. The reason
is that in Ref.~\cite{el03}, redshift distortions are included in
the 2dFGRS data analysis. Given a constraint on the amplitude of fluctuations
from CMB data, and a constraint on $\Omega_m h^2$ , this effectively
constrains the bias parameter. Therefore adding a further constraint
on bias in their analysis does not change the results.

\begin{figure}[htb]
\begin{center}
\includegraphics[width=.8\textwidth]{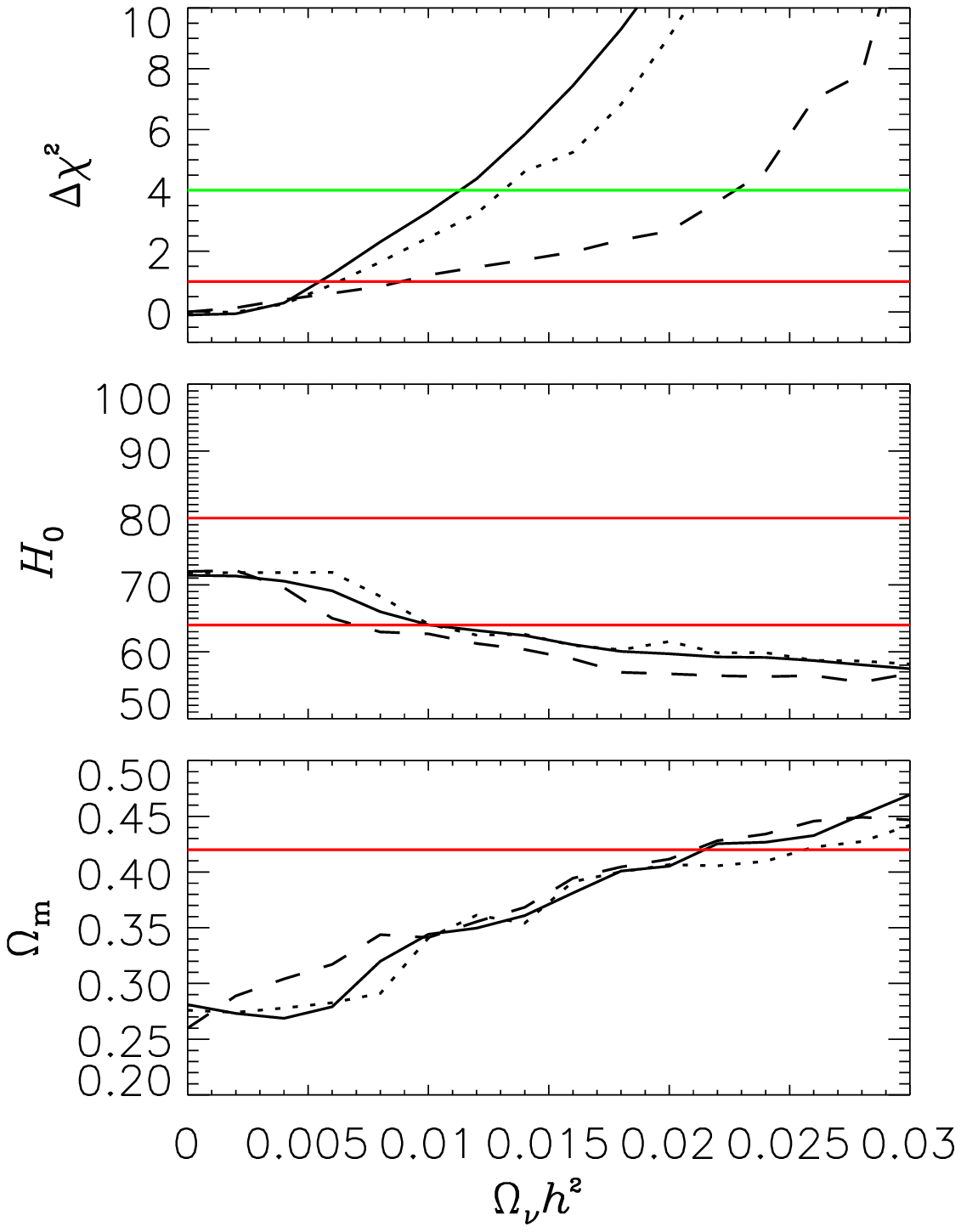}
\end{center}
\caption[]{The top panel shows
$\chi^2$ as a function of $\sum m_\nu$ for different choices
of priors. The dotted line is for WMAP + 2dFGRS data alone,
the dashed line is with the additional Wang {\it et al.} data.
The full line is for additional HST and SNI-a
priors as discussed in the text. The horizontal lines show
$\Delta \chi^2 = 1$ and 4 respectively. 
The middle panel shows the best fit values of $H_0$ for a given
$\sum m_\nu$. The horizontal lines show the HST key project
$1\sigma$ limit of $H_0 = 72\pm 8 \,\, {\rm km}\,{\rm s}\,{\rm Mpc}^{-1}$.
Finally, the lower panel shows best fit values of $\Omega_m$. In this
case the horizontal line corresponds to the SNI-a $1\sigma$ upper limit
of $\Omega_m < 0.42$.}
\label{fig1}
\end{figure}

{\it Neutrinoless double beta decay}

Recently it was claimed that the Heidelberg-Moscow experiment
yields positive evidence for neutrinoless double beta decay.
Such experiments probe the `effective electron neutrino mass
$m_{ee} = |\sum_j U^2_{ej} m_{\nu_j}|$. Given the uncertainties in
the involved nuclear matrix elements the Heidelberg-Moscow
result leads to a mass of $m_{ee} = 0.3-1.4$ eV. If this is
true then the mass eigenstates are necessarily degenerate, and
$\sum m_\nu \simeq 3 m_{ee}$.
Taking the WMAP result of $\sum m_\nu \leq 0.70$ eV at face
value seems to 
be inconsistent with the Heidelberg-Moscow result
\cite{Pierce:2003uh}.
However, already if Ly-$\alpha$ forest data and a constraint
on the bias parameter is not used in the analysis
the upper bound of $\sum m_\nu \leq 1.01$ eV is still
consistent.
For this reason it is probably premature to rule out the
claimed evidence for neutrinoless double beta decay.


\subsection{Neutrino relativistic energy density}

For the case of the effective number of neutrino species we have
in all cases calculated constraints in the $(\Omega_b h^2, N_\nu)$ 
plane, while marginalizing over all other parameters. 
The reason for this is that for Big Bang Nucleosynthesis
purposes these are exactly the
important parameters. Therefore, to combine CMB, LSS, and BBN constraints
the marginalization over $\Omega_b h^2$ should not be performed.
Furthermore, 
when constraining $N_\nu$ we have always assumed that $\sum m_\nu \simeq 0$
so that the neutrino mass has no influence on cosmology.

We start out by investigating constraints on $N_\nu$ from CMB and LSS
data alone. In Fig.~2 we show $\Delta \chi^2$ for a global fit to
$N_\nu$ which marginalizes over all other cosmological parameters.
The overall best fit for the WMAP $T$ and $TE$ data, combined with
the Wang {\it et al.} compilation, the 
2dFGRS data, the HST key project data on $H_0$, as well
as the SNI-a data on $\Omega_m$, has $\chi^2 = 1467.6$ for
a total of 1395 degrees of freedom. This gives $\chi^2/{\rm d.o.f.}
= 1.05$ which is entirely compatible with the best fit WMAP
value for the standard $\Lambda$CDM model of $\chi^2/{\rm d.o.f.}
= 1.066$. We also show constraints for two other analyses. The first
is for WMAP and 2dFGRS data alone and the second for WMAP data alone.
The bounds for the three cases are
\begin{tabbing}
$N_\nu = 4.0^{+3.0}_{-2.1}$ \hspace*{0.8cm} \= for WMAP+Wang+2dFGRS+HST+SN-Ia \\
$N_\nu = 3.1^{+3.9}_{-2.8}$ \> for WMAP+2dFGRS \\
$N_\nu = 2.1^{+6.7}_{-2.2}$ \> for WMAP only
\end{tabbing}
These bound are entirely compatible with those found by 
Crotty, Lesgourgues and Pastor \cite{CLP}, and much
tighter than the pre-WMAP constraints.

The constraints derived here are also compatible with what is found
by Pierpaoli \cite{pierpaoli}, where are assumption of spatial flatness
was relaxed.

In the lower panels of Fig.~2 we show the best fit values of $H_0$
and $\Omega_m$ for a given value of $N_\nu$.
The main point to note is that the constraint on $N_\nu$ is strongly
dependent on $H_0$. This was also found in Ref.~\cite{Hannestad:2001hn}.
With only CMB data and the weak top-hat prior on $H_0$ the bound
on $N_\nu$ is very weak. Adding the HST Key Project prior on
$H_0$ cut away a significant amount of parameter space both at
low and high $N_\nu$. Adding the 2dFGRS and Wang {\it et al.} data
mainly has the effect of shifting the best fit value to higher $N_\nu$,
but also cuts away the low $N_\nu$ values, an effect also seen in
Ref.~\cite{CLP}.

\begin{figure}[htb]
\begin{center}
\includegraphics[width=.8\textwidth]{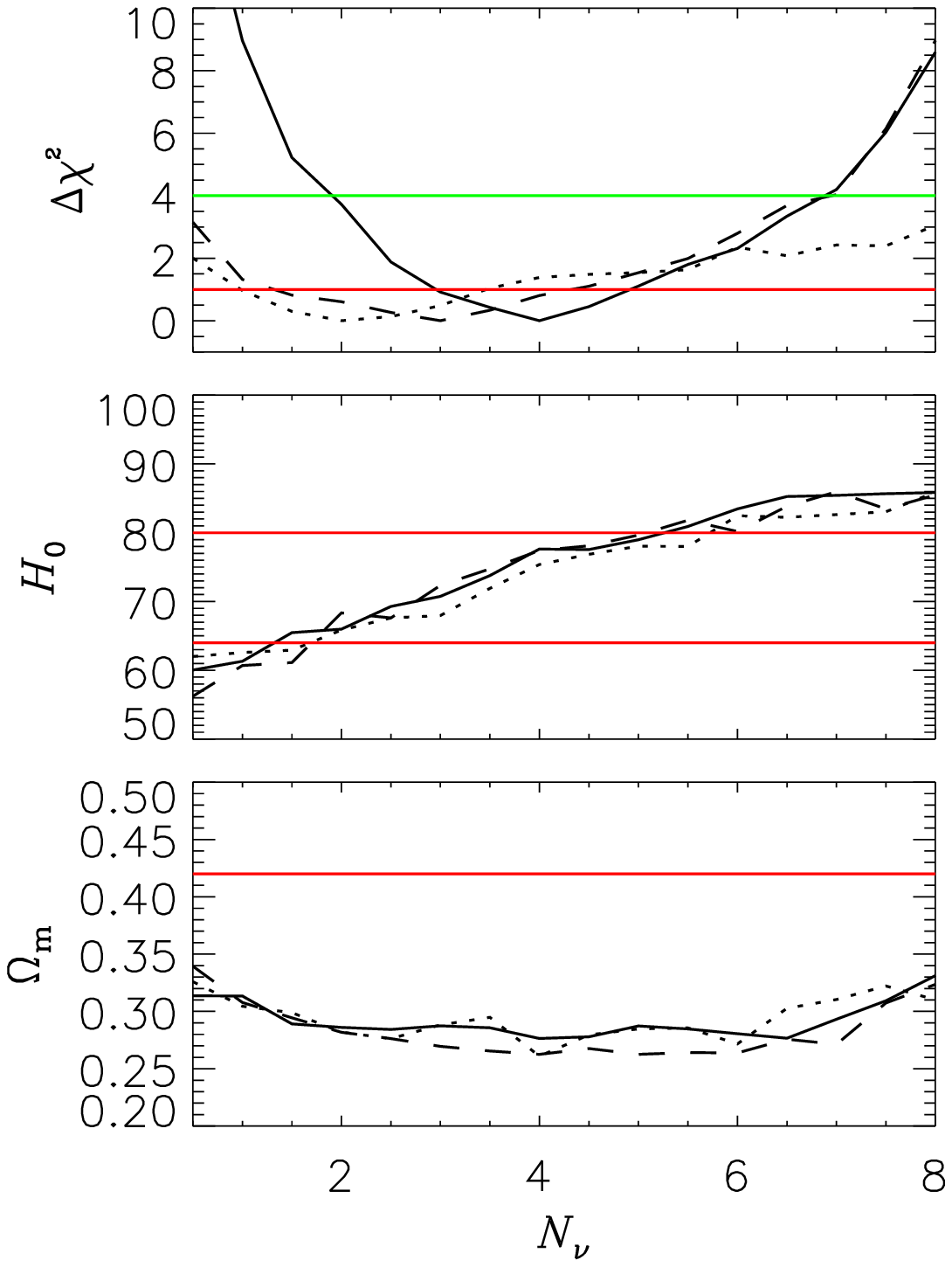}
\end{center}
\caption[]{$\chi^2$ as a function of $\sum N_\nu$ for different choices
of priors. The dotted line is for WMAP data alone,
the dashed line is with the additional Wang {\it et al.} and 2dFGRS data.
The full line is for additional HST and SNI-a
priors as discussed in the text.The horizontal lines show
$\Delta \chi^2 = 1$ and 4 respectively. 
The middle panel shows the best fit values of $H_0$ for a given
$N_\nu$. The horizontal lines show the HST key project
$1\sigma$ limit of $H_0 = 72\pm 8 \,\, {\rm km}\,{\rm s}\,{\rm Mpc}^{-1}$.
Finally, the lower panel shows best fit values of $\Omega_m$. In this
case the horizontal line corresponds to the SNI-a $1\sigma$ upper limit
of $\Omega_m < 0.42$.}
\label{fig2}
\end{figure}

In Fig.~3 we show constraints on $(\Omega_b h^2, N_\nu)$ for the
full data set described above. The best fit value for 
$\Omega_b h^2$ is 0.0233, which is equivalent to the value found
in the WMAP data analysis. In the 2-dimensional plots the
68\% and 95\% regions are formally
defined by $\Delta \chi^2 = 2.30$ and 6.17
respectively. Note that this means that the 68\% and 95\% contours
are not necessarily equivalent to the same confidence level for single
parameter estimates.

\begin{figure}[htb]
\begin{center}
\includegraphics[width=.8\textwidth]{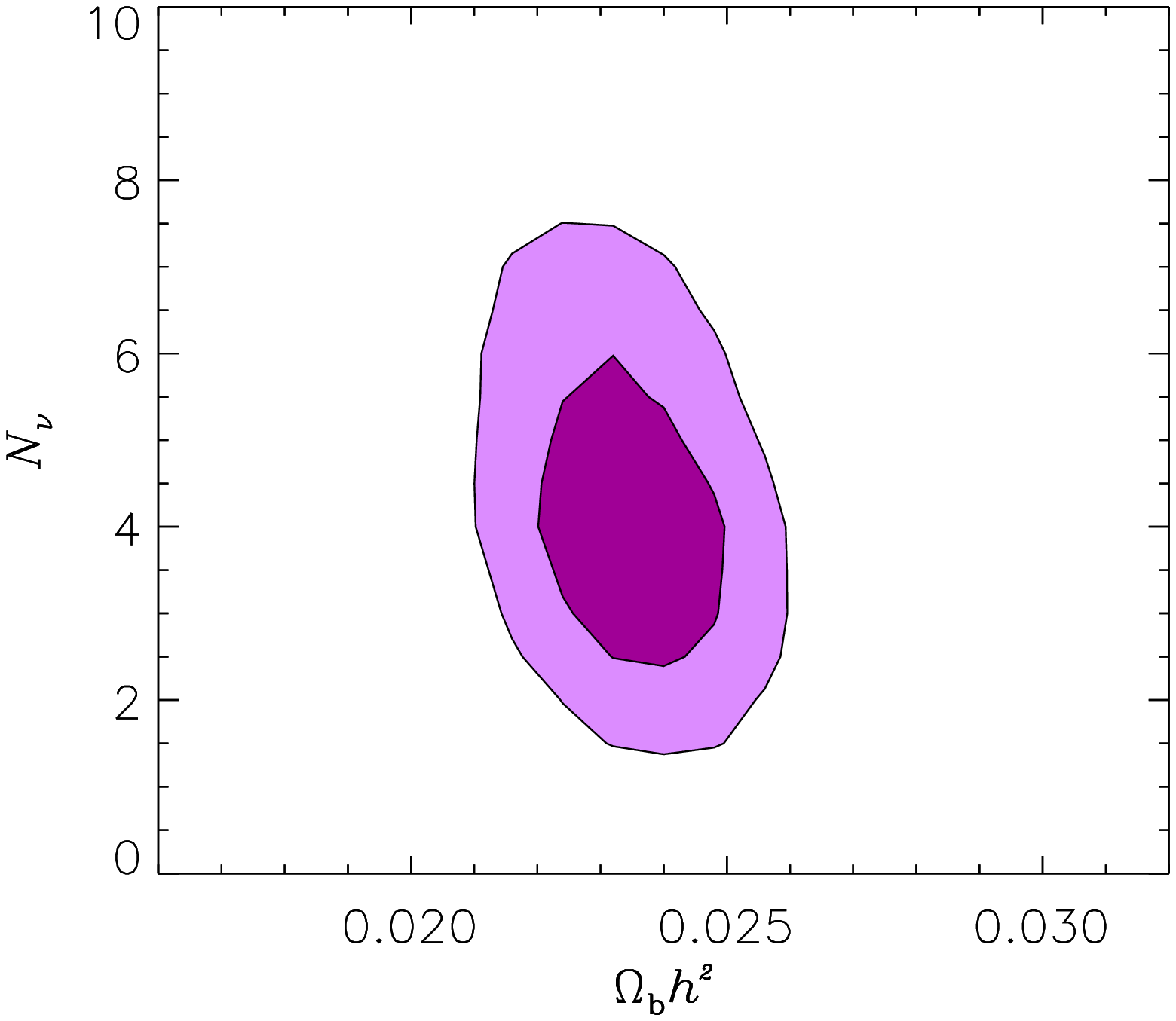}
\end{center}
\caption[]{68\% and 95\% confidence contours in the $(\Omega_b h^2,N_\nu)$
plane for the WMAP TT and TE data, combined with the 2dFGRS data, the
HST data on $H_0$ and the SNI-a data on $\Omega_m$.}
\label{fig3}
\end{figure}

It should be noted here that in addition to an upper bound on $N_\nu$
there is also a 3.0$\sigma$ confidence detection of $N_\nu > 0$.
This is in concordance with the pre-WMAP data from which a non-trivial
lower bound on $N_\nu$ could also be derived.

{\it Adding BBN information --} In the case where all the relativistic 
energy density contained in $N_\nu$ is produced prior to BBN, a
BBN constraint can be added to the CMB and LSS constraint without
any problems. In practice we have used abundances of He-4 and D
to make constraints in the $(\Omega_b h^2,N_\nu)$ plane. We use
the following values for the primordial abundances
\cite{kirkman,steigman}
${\rm D/H}  =  2.78^{+0.44}_{-0.38} \times 10^{-5}$ and
${\rm Y}_P  =  0.238 \pm 0.005$

This calculation is shown in Fig.~4. In terms of a single parameter
constraint on $N_\nu$ it is $N_\nu = 2.6^{+0.4}_{-0.3}$ (95 \% conf.).
Compared to the recent calculation by Abazajian \cite{Abazajian:2002bj}
of a BBN-only constraint
of $1.7 \leq N_\nu \leq 3.5$ (95 \% conf.) this is a significant improvement.
Very interestingly the new limit suggests the possibility that 
$N_\nu$ is actually less than 3. This is for instance possible in
scenarios with extremely low reheating temperature
\cite{Giudice:2000ex,Kawasaki:2000en}.
Note that this result is also consistent with the calculation
by Barger et al. \cite{Barger:2003zg}, although t
heir preferred region is slightly
different because they do not include large scale structure data in their
analysis.

Of course this conclusion is mainly based on the fact that CMB and LSS
data prefers a slightly higher value of $\Omega_b h^2$ than BBN.
It should also be stressed that the estimates of the
primordial abundances could be biased by systematic effects so that the
quoted statistical error bar is not really meaningful.
Therefore
it is probably premature to talk of any inconsistency between the 
$N_\nu = 3$ prediction of the standard model and observations.

In fact the argument can also be reversed. If $N_\nu$ is fixed to the
standard model value of 3 then then CMB and LSS constraint on $\Omega_b h^2$
provides an accurate measurement of primordial He-4. Using the derived
constraint on $\Omega_b h^2$ the 95\% confidence range for $Y_P$ is
$0.2458 \leq Y_P \leq 0.2471$.
This could point to a serious underlying systematic effect in 
observational determinations of $Y_P$, as discussed in 
Ref.~\cite{Cyburt:2003fe}.

\begin{figure}[htb]
\begin{center}
\includegraphics[width=.8\textwidth]{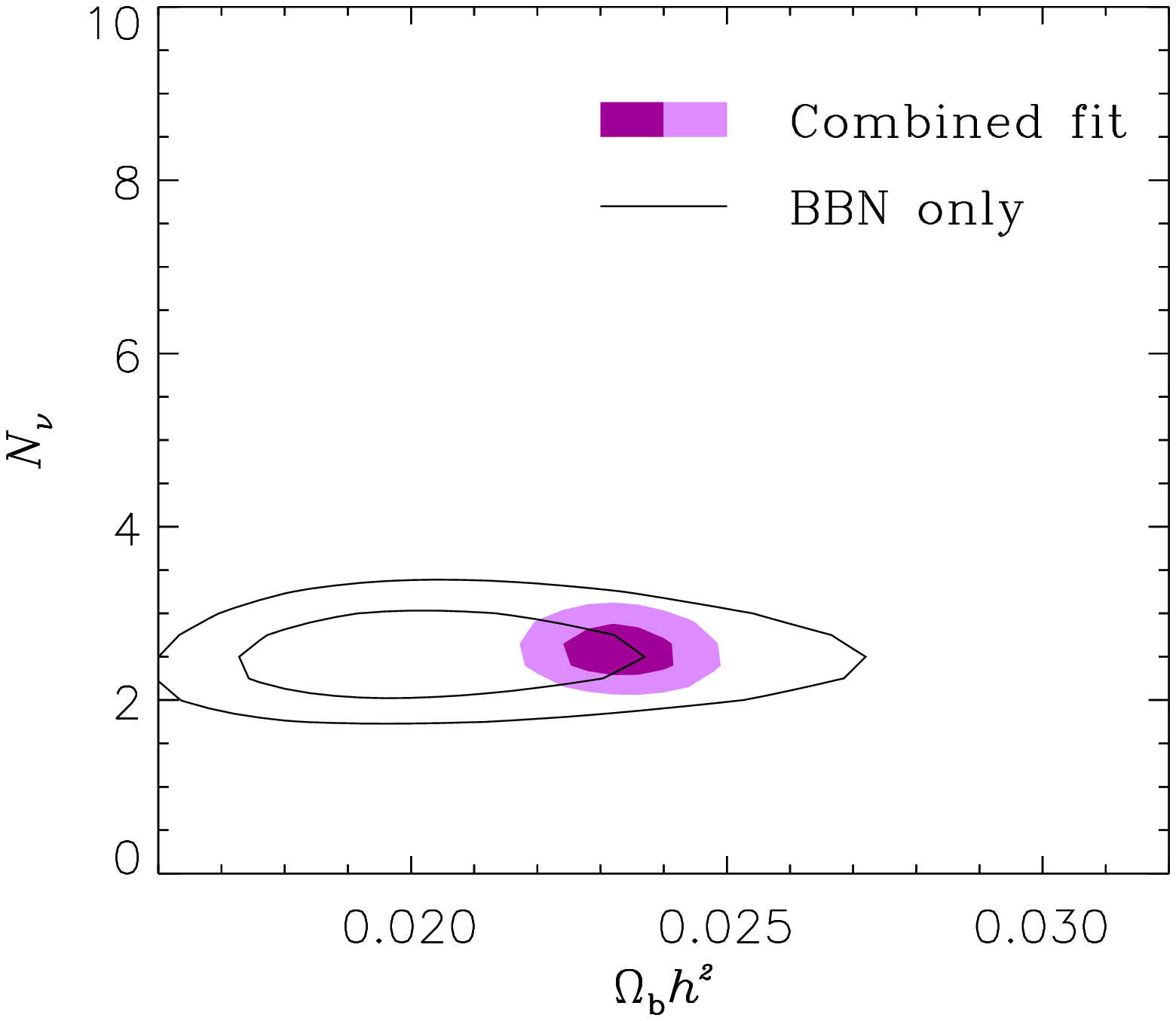}
\end{center}
\caption[]{68\% and 95\% confidence contours in the $(\Omega_b h^2,N_\nu)$
plane for the same data sets as in fig.~3, but with the addition of BBN
data. The lined contours are the 68\% and 95\% regions for BBN data alone.}
\label{fig4}
\end{figure}

\subsection{Combining $\sum m_\nu$ with $N_\nu$ - constraining LSND}

From the analyses in the above two sections it was found that:
{\it (a)} An increasing $\sum m_\nu$ can be compensated by a decreasing
$H_0$ and {\it (b)} An increasing $N_\nu$ can be compensated by
an increasing $H_0$. One might therefore wonder whether a model
with non-zero $\sum m_\nu$, combined with $N_\nu > 3$ can provide
a good fit to the data.

In order to test this we have performed a likelihood analysis
for $\sum m_\nu$ for different values of $N_\nu$. We show this
in Fig.~5. This analysis was performed with all available data and 
priors. 

As can be seen from the figure, the best fit actually is actually 
shifted to higher $\sum m_\nu$ when $N_\nu$ increases, and the conclusion
is that a model with high neutrino mass and additional relativistic
energy density can provide acceptable fits to the data. As a function
of $N_\nu$ the upper bound on $\sum m_\nu$ is (at 95\% confidence)
\begin{tabbing}
$\sum m_\nu \leq 1.01$ eV \hspace*{0.8cm} \= for $N_\nu = 3$ \\
$\sum m_\nu \leq 1.38$ eV \hspace*{0.8cm} \= for $N_\nu = 4$ \\
$\sum m_\nu \leq 2.12$ eV \hspace*{0.8cm} \= for $N_\nu = 5$
\end{tabbing}

\begin{figure}[htb]
\begin{center}
\includegraphics[width=.8\textwidth]{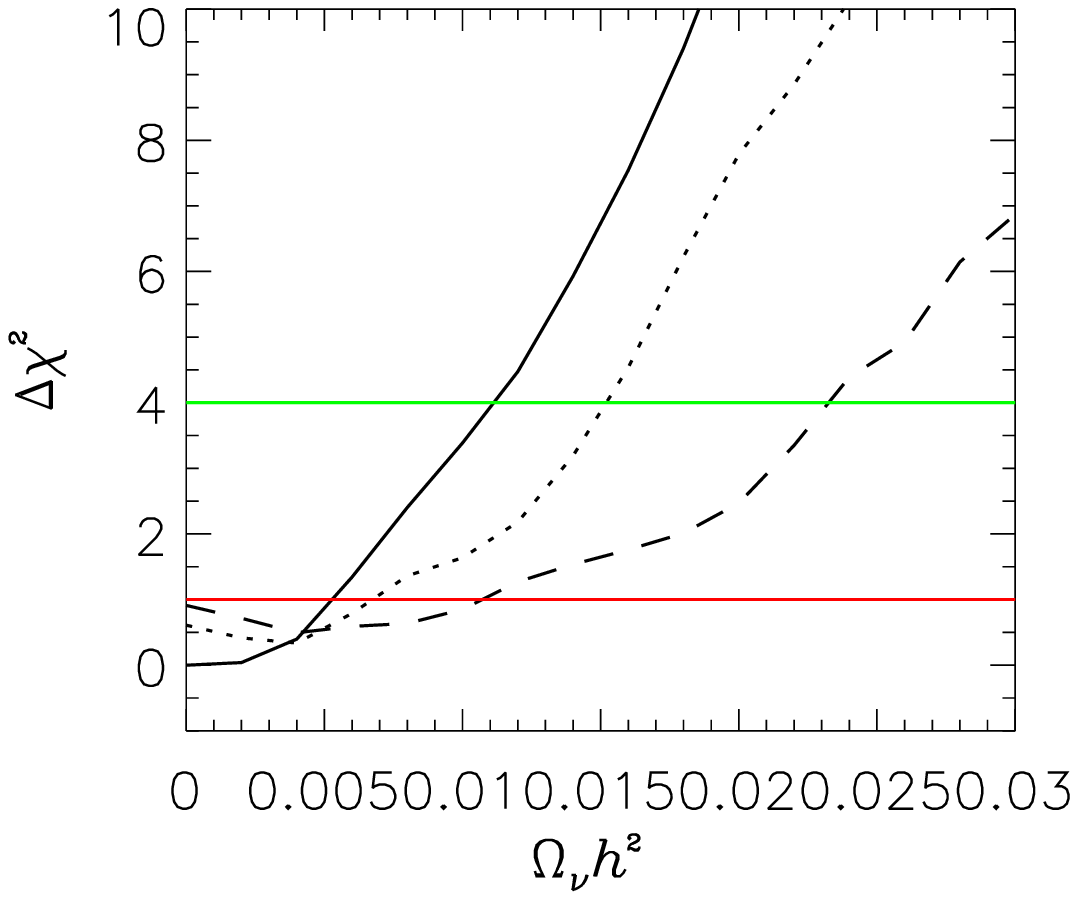}
\end{center}
\caption[]{$\Delta \chi^2$ as a function of $\sum m_\nu$ for various
different values of $N_\nu$. The full line is for $N_\nu = 3$, the dotted
for $N_\nu = 4$, and the dashed for $N_\nu = 5$. $\Delta \chi^2$ is calculated
relative to the best fit $N_\nu = 3$ model.}
\label{fig5}
\end{figure}

This has significant implications for attempts to constrain the
LSND experiment using the present cosmological data.
Pierce and Murayama conclude from the present MAP
limit that the LSND result is excluded \cite{Pierce:2003uh} 
(see also Ref.~\cite{Giunti:2003cf}).

However, for several reasons this conclusion does not follow
trivially from the present data. 
In general the three mass differences implied by Solar, atmospheric
and the LSND neutrino measurements can be arranged into either 
2+2 or 3+1 schemes.
Recent analyses \cite{Maltoni:2002xd}
of experimental data have shown that the 2+2 
models are ruled out. The 3+1 scheme with a single massive
state, $m_4$, which makes up the LSND mass gap, is still
marginally allowed in a few small windows in the 
$(\Delta m^2,\sin^2 2 \theta)$ plane. These gaps are at 
$(\Delta m^2,\sin^2 2 \theta) \simeq  (0.8 \, {\rm eV}^2, 2 \times 10^{-3}),
(1.8 \, {\rm eV}^2, 8 \times 10^{-4}), 
(6 \, {\rm eV}^2, 1.5 \times 10^{-3})$ and
 $(10 \, {\rm eV}^2, 1.5 \times 10^{-3})$.
These four windows corresponds to masses of $0.9, 1.4, 2.5$ and 3.2 eV
respectively.
From the Solar and atmospheric neutrino results the three light
mass eigenstates contribute only about 0.1 eV of mass if they are
hierarchical. This means that the sum of all mass eigenstate is close
to $m_4$.

The limit for $N_\nu = 4$ which corresponds roughly to the LSND
scenario is $\sum m_\nu \leq 1.4$ eV, which still leaves
the lowest of the remaining windows. The second window at $m \sim 1.8$
eV is disfavoured by the data, but not at very high significance.


\section{Discussion}

We have calculated improved constraints on neutrino masses
and the cosmological relativistic
energy density, using the new WMAP data together with data from the
2dFGRS galaxy survey.

Using CMB and LSS data together with a prior from the HST key project
on $H_0$ yielded an upper bound of $\sum m_\nu \leq 1.01$ eV
(95\% conf.). While this excludes most of the parameter range
suggested by the claimed evidence for neutrinoless double
beta decay in the Heidelberg-Moscow experiment, it seems premature
to rule out this claim based on cosmological observations.

Another issue where the cosmological upper bound on neutrino
masses is very important is for the prospects of directly measuring
neutrino masses in tritium endpoint measurements.
The successor to the Mainz experiment, KATRIN, is designed to
measure an electron neutrino mass of roughly 0.2 eV,
or in terms 
of the sum of neutrino mass eigenstates, $\sum m_\nu \leq 0.75$ eV.
The WMAP result of $\sum m_\nu \leq 0.7$ eV (95\% conf.) already
seems to exclude a positive measurement of mass in KATRIN.
However, this very tight limit depends on priors, as well
as Ly-$\alpha$ forest data, and the more conservative present
limit of $\sum m_\nu \leq 1.01$ eV (95\% conf.) does not exclude
that KATRIN will detect a neutrino mass.

From the data we also inferred a limit on $N_\nu$ of 
$N_\nu = 4.0^{+3.0}_{-2.1}$ (95\% conf.)
on the equivalent number of neutrino species.
This is a marked improvement over the previous best limit of roughly
$N_\nu \leq 13$ \cite{Hannestad:2001hn,Hannestad:2000hc}.

When light element measurements of He-4 and D are included
the bound is strengthened considerably to $N_\nu = 2.6^{+0.4}_{-0.3}$
(95 \% conf.). Interestingly this suggests a possible value of
$N_\nu$ which is {\it less} than 3. This could be the case for
instance in scenarios with very low reheating temperature where neutrinos
were never fully equilibrated 
\cite{Giudice:2000ex,Kawasaki:2000en}.
However, it should be stressed that primordial abundances could be
dominated by systematics. Therefore it is probably premature
to talk of a new BBN ``crisis''.

Finally, we also found that the neutrino mass bound depends on
the total number of light neutrino species. In scenarios with sterile
neutrinos this is an important factor. For instance in 3+1 models
the mass bound increases from 1.0 eV to 1.4 eV, meaning that the
LSND result is not ruled out by cosmological observations yet.

\end{document}